\documentstyle[12pt,aaspp4]{article}
\begin{document}
\title{Cosmology with Modified Newtonian Dynamics (MOND)}
\author{R.H. Sanders}
\affil{Kapteyn Astronomical Institute, Groningen, The Netherlands}

\begin{abstract}

It is well-known that the application of Newtonian
dynamics to an expanding spherical region leads to the correct relativistic
expression (the Friedmann equation) for the evolution of the cosmic scale 
factor.  Here, the cosmological implications of Milgrom's modified 
Newtonian dynamics (MOND) are considered by means of a similar procedure.
Earlier work by Felten demonstrated that in a region dominated by 
modified dynamics, the expansion cannot be uniform (separations cannot
be expressed in terms of a scale factor) and that any such region will
eventually re-collapse regardless of the initial expansion velocity and 
mean density.  Here I show that, because of the acceleration threshold
for the MOND phenomenology, a region dominated by MOND 
will have a finite size which, in the early Universe ($z > 3$), is smaller 
than the horizon scale.  Therefore, uniform expansion and homogeneity
on horizon scale are consistent with MOND-dominated non-uniform 
expansion and the development of inhomogeneities on 
smaller scale.  In the radiation-dominated era, the amplitude of MOND-induced 
inhomogeneities is much smaller than that implied by observations of
the cosmic background radiation, and the thermal and dynamical history
of the Universe is identical to that of the standard Big Bang.  
In particular, the standard
results for primordial nucleosynthesis are retained.  When matter first 
dominates the energy density of the Universe, the cosmology diverges
from that of the standard model.  Objects of galaxy mass are the first
virialized objects to form (by z=10) and larger structure develops rapidly.
At present, the Universe would be inhomogeneous out to a substantial
fraction of the Hubble radius.

\end{abstract}

\section{Introduction}

The modified Newtonian dynamics (MOND) proposes that the law of inertia
or gravity takes on a specific non-Newtonian form at accelerations well 
below a definite universal value (Milgrom 1983a,b,c).  As an 
alternative to dark
matter, the simple MOND formula with one new fixed parameter (the critical 
acceleration $a_o$) has been quite successful in predicting the form
of spiral galaxy rotation curves from the observed distribution of 
detectable matter (Begeman et al. 1990, Sanders 1996) as well as the 
magnitude
of the conventional mass discrepancy in galaxy clusters (Milgrom 1983c) and
in superclusters (Milgrom 1997).  MOND subsumes the global
scaling relationships for galaxies-- the Tully-Fisher relation for
spirals and the Faber-Jackson relation for ellipticals, as well as an
equivalent gas temperature-mass relation for clusters of galaxies
(Sanders 1994).  MOND stabilizes rotationally supported thin disks
(Brada 1996),
explains the presence of a maximum critical surface density in 
spiral galaxies and ellipticals (Milgrom 1983b), 
and predicts the observed large
conventional mass discrepancy in low-surface brightness systems
(Milgrom 1983b, McGaugh \& de Blok 1997).

However, an argument often directed against MOND is that, as a theory,
it is ad hoc and incomplete.  In particular, MOND makes no predictions
with respect to cosmology and cosmogony.  Even though the 
near coincidence of
the empirically determined $a_o$ with $cH_o$ is suggestive of a 
cosmological basis for MOND, the structure of that cosmology is not
at all evident.  The naive expectation is that
a hypothesis which posits such a radical departure from Newtonian dynamics
(and presumably General Relativity) on the scale of galaxies and clusters
of galaxies would surely lead to a highly unconventional cosmology which
would be inconsistent with the phenomenological successes of the
standard Big Bang-- most notably the nucleosynthesis of the light
elements in their observed abundances and the large and small scale isotropy
of the Universe at the epoch of recombination.  Such criticism 
cannot be addressed by an incomplete theory. 

The reason for this incompleteness is that
MOND at the present time lacks a relativistic extention; there is no
credible general theory of gravity which predicts the MOND phenomenology
in the appropriate limit.  In fact, non-standard scalar-tensor theories have
been proposed as a theoretical underpinning of MOND (Bekenstein and
Milgrom 1984, Bekenstein 1988, Romatka 1992, Sanders 1997).  Two of these, 
phase-coupling gravity (Bekenstein 1988) and stratified (preferred frame)
theories with aquadratic scalar field Lagrangians (Bekenstein \& Milgrom
1984, Sanders 1997) do yield sensible cosmologies--
isotropic and homogeneous and similar to the low-density Friedmann
models (Sanders 1989, Sanders 1997).  Although such theories do have
the considerable advantage of predictive power on scales other than
extra-galactic, the fact is that they are unnatural;  these non-standard
scalar-tensor theories are as contrived as the ad hoc modification of
Newton's laws which they presume to replace.  Thus it is perhaps 
premature to work out fully the cosmological implications of such
complicated and tentative theories.  

Must, then, considerations of
MOND cosmology then be postponed until the final theory is in place?
Even before further theoretical developments, it may be
possible to draw some preliminary conclusions about a
MOND universe by considering a finite expanding spherical region.
It is well-known that the application of Newtonian dynamics to such 
a region leads to the Friedmann equation for the evolution of the
cosmic scale factor.  Even the cosmological constant can be included
as an additional fluid with negative pressure.
It might be expected that some insight into a pure MOND cosmology might
be gained by such a procedure using Milgrom's formula instead of 
Newton's.

An interesting start in this direction was made by Felten (1984).  He
pointed out that with MOND, the physical size of the expanding region
cannot be factored out.  This means that uniform expansion of a spherical
region is not possible;  that the dynamical history of such a region 
in the Universe depends
upon its physical size.  This would suggest that an isotropic and
homogeneous universe, as described by the Robertson-Walker (RW)
metric is not
possible in the context of MOND.  Moreover, due to the effective 
logarithmic potential implied by MOND, any finite-size region will 
re-collapse in a finite time.  The universe, out to the present horizon,
will eventually re-collapse regardless of its mean density.  In a 
low-density Universe a region 
with a present size of 20 to 30 Mpc would just now be turning around,
and this, as stressed by Felten, is roughly the observed
scale of large-scale structure--  voids and superclusters.

In the present paper I continue the discussion of Felten on MOND 
cosmology in the context of finite expanding spherical regions.
Three assumptions underly this discussion, all of which were 
more or less implicit
in the work of Felten.  The first is that the dynamics of such a
region are not influenced by the exterior universe--
that there is, in effect, an equivalent of the Birkhoff theorem for
the relativistic theory of MOND.  This assumption is the most shaky.
Scalar-tensor theories of modified dynamics violate the strong 
equivalence principle which means that no dynamical system is isolated
from its environment.  The time-variation of the gravitational constant
due to universal expansion is one example of the possible effect of the 
rest of the Universe on the spherical region.  Here the assumption is that
any such effects which may be present in the final theory 
will play a minor role in the dynamical history of the finite volume.

The second assumption is that the modified dynamics holds for all 
accelerations below $a_o$-- that there is no return to Newtonian
attraction or inertia at very much lower accelerations, as in 
cosmological PCG.  This then will be an exploration of the cosmology
of pure MOND.  

Finally, it is assumed that the critical acceleration $a_o$ is independent
of cosmic time (not true in cosmological PCG).  The fact
that $a_o\approx cH_o$ would suggest that $a_o$ is time dependent
(as is the Hubble parameter).  However, this expression could have one
of several meanings as pointed out by Milgrom (1994).  One possible basis
is that $a_o\approx c{\sqrt{\Lambda}}$ where $\Lambda \approx {H_o}^2$
is the cosmological constant.  Such an interpretation would consistent
with the assumption of no time variation of $a_o$.  Alternatively, the
numerical coincidence could arise from anthropic considerations:  as
will be shown below, structure develops when $cH\rightarrow a_o$.

Having made these assumptions, I re-derive Felten's expression for the 
evolution of an isolated spherical region dominated by non-relativistic,
pressure-less matter.  I then demonstrate that, because of the
acceleration threshold, modified dynamics can only be valid inside
a critical radius $r_c$.  This critical radius expands faster than the
scale factor and the horizon, so that in the earlier universe
the size of the region in which MOND applies is 
smaller than the horizon scale.  

This is the principal result of the present paper.  Friedmann cosmology, 
characterized by uniform expansion, applies on the scale of the
horizon, while
MOND, in which the expansion is highly non-uniform,
applies on sub-horizon scales.  Therefore, the usual 
Friedmann equation is valid for evolution of the universal scale factor,
but the Felten equation is valid for spherical regions smaller than
$r_c$.  This means that, at any epoch, while the Universe overall is
homogeneous, density 
inhomogeneities should be present on the scale of $r_c$ at that time.  
Only recently in the history of the Universe has the size of the
region dominated by MOND expanded to include a significant fraction of the
the observable Universe, c/H$_o$ ($r_c$ at present depends upon
the value of the cosmological constant); 
i.e., the Universe on large scale has only become ``MONDIAN'' at late
cosmic time.  

These results remain qualitatively the same when radiation is included.
In the early universe, at the epoch of nucleosynthesis, MOND regions are
very much smaller than the horizon.  When pressure gradient forces are 
considered, the density and 
expansion of the universe remain highly uniform during the 
radiation-dominated 
era and identical to that of the standard hot Big Bang,  so all results
concerning primordial nucleosynthesis are retained.  Moreover, 
MOND-induced inhomogeneities at the epoch of recombination are many
orders of magnitude smaller than those implied by the observed 
fluctuations in the background radiation.

Regions of larger comoving size and mass enter the regime of modified 
dynamics at later times.
Re-collapse of finite size regions dominated by modified dynamics proceeds
rapidly once non-relativistic matter dominates the energy density,
even in a very low density universe.  This is due to the effective
logarithmic gravitational potential implied by MOND. 
At the epoch of matter-radiation equality the mass enclosed within a 
MOND dominated region is $\approx 10^9$ M$_\odot$, comparable to a 
low mass galaxy.  This gives a preferred mass scale to the first objects
which collapse and virialize and may explain why galaxy mass objects
are the fundamental virialized building blocks.

The expansion of MOND-dominated regions to
include larger and larger comoving scales leads to a scenario of structure 
formation which is extremely hierarchical and ``bottom up'', 
with the smallest objects
forming first-- star clusters and low mass galaxies-- 
and larger structures forming
later by a series of mergers.  Massive galaxies should be in place 
as virialized
objects by a redshift of 5 to 10.  The largest objects just now being 
virialized are the rich clusters and supercluster scale objects would only
now be separating out of Hubble expansion.  The scale-dependent 
deceleration induced by MOND implies that the Universe, at the present
epoch, is inhomogeneous
on large scale with a mean density of galaxies about a given galaxy
decreasing out to hundreds of Mpc.  The scale for cross-over to homogeneity
depends the value of the cosmological constant.

\section{Dynamics of an Isolated Spherical Region}

The Cosmological Principle which postulates the isotropy and homogeneity
of the Universe permits separations between physical objects
to be described by a universal
dimensionless scale factor which is only a function of cosmic time.
It is well-known (eg. see Peebles 1993) that the Friedmann equation for the 
evolution of scale factor can be derived by considering the Newtonian
equation of motion for an isolated uniform spherical region of radius r:
$$\ddot{r} = -{{GM}\over {r^2}}. \eqno(1)$$
Here M is the active gravitational mass given, in the weak-field static
limit of the Einstein field equations, by
$$M = {{4\pi r^3}\over 3}(\rho + 3p)\eqno(2)$$
where $\rho$ and $p$ are the density and pressure of the smooth fluid.
Combining eqs. 1 and 2 we have
$$\ddot{r} = - {{4\pi G r}\over 3}{(\rho + 3p)} \eqno(3)$$
which is supplemented by the energy conservation equation
$${{d\rho}\over{(\rho+p)}} = -{{3dr}\over r}\eqno(4)$$
and an equation of state.  Obviously one may write $r = Lx$
where L is a fixed length scale and x is a time-dependent scale factor
(here normalized to be 1 at the present epoch).
Then L disappears in eq.\ 3 and 4.
Taking the fluid to be a mixture of non-relativistic pressure-less matter
($p=0$), radiation ($p_r = {1\over3} \rho_r$), and vacuum energy density
($p_v = -\rho_v = -{{3\lambda {H_o}^2}/8\pi G}$)
and integrating eq.\ 3 we find the usual dimensionless Friedmann equation
for the time-evolution of the scale factor,
$$ h^2=\Bigl({{\dot{x}\over x}\Bigr)^2 = {\Omega_o}x^{-3}} + {\Omega_r}x^{-4}
- (\Omega_o +\Omega_r+\lambda-1)x^{-2} + \lambda \eqno(5)$$
where $\lambda$ is the dimensionless cosmological constant,
$$\Omega_o = {{8\pi G \rho_o}\over{3{H_o}^2}}\eqno(6)$$
is the density parameter for non-relativistic matter (with $\rho_o$
being the present mean density of matter) and
$$\Omega_r = {{8\pi G a{T_o}^2}\over {3{H_o}^2}c^2}\eqno(7)$$
is the density parameter for the cosmic background radiation
where $T_o$ is the temperature of the cosmic blackbody radiation at the
present epoch (2.73 K) and $a$ is the radiation density constant.  
The quantity $(\Omega_o+\Omega_r + \lambda -1)$
is the integration constant
which is to be identified with curvature of space-time.
The quantity $h$ is the Hubble parameter in units of the
present Hubble parameter $H_o$ and time is in units of the Hubble time
$\rm{{H_o}^{-1}}$.

MOND posits that for accelerations below a critical value $a_o$, the
true gravitational force $g$ is related to the Newtonian gravitation
force $g_n$ as
$${\bf g}\mu(g/a_o) = {\bf g_n} \eqno(8)$$
(Milgrom 1983a) where $\mu(x)$ is an unspecified function such that
$\mu(x)\rightarrow 1$ if $x>>1$ and $\mu(x)=x$ if $x<<1$.
Thus in the high acceleration limit the gravitational force is the 
usual Newtonian force, but in the low acceleration limit
$g = \sqrt{g_n a_o}$ (this may also be written as a modification of the
law of inertia where $F = ma\mu(a/a_o)$ replaces the usual expression).
Because we are interested here only in a broad view of the overall
dynamics of a MOND universe, we will assume that the transition in
$\mu(x)$ between the two asymptotic limits occurs abruptly at x=1.

In the low acceleration limit the MOND equivalent of eq.\ 3 becomes
$$\ddot{r} = -\Bigl[{{4\pi G a_o}\over 3}(\rho+3p)r\Bigr]^{1\over 2}\eqno(9)$$
where now, obviously, a constant length scale cannot be factored out.
Neglecting for the moment radiation and vacuum energy density and
taking the equation of state only for non-relativistic pressure-less matter,
the conservation equation implies that
$$\rho = \rho_o (r/r_o)^{-3}\eqno(10)$$
where $r_o$ is the comoving radius of the spherical region (i.e., the radius
the spherical region would have at present if it continued to expand
according to eq.\ 5) and $\rho_o$ is the present mean density in the 
equivalent Friedmann model universe.  Then eq.\ 9 becomes
$$\ddot{r} = -\Bigl[{{\Omega_o}\over 2}{H_o}^2{r_o}^3 a_o\Bigr]^{1\over2} 
r^{-1}. \eqno(11)$$
This equation may be integrated once to give the equivalent of the 
Friedmann equation
$${\dot{r}}^2 = {u_i}^2 - \Bigl[{{2\Omega_o}}{H_o}^2{r_o}^3 a_o
\Bigr]^{1\over 2} ln(r/r_i)\eqno(12)$$
where $r_i$ is an initial radius of the sphere
and $u_i$ is the expansion 
velocity at this initial radius.  From the form of eq.\ 12 it is 
obvious that at some maximum radius $r_m$ the expansion will stop and
the spherical region will re-collapse.  This is given by
$$r_m/r_i = e^{q^2} \eqno(13a)$$
where $$q^2 = {{{u_i}^2}\over{{(2\Omega_o{H_o}^2}{r_o}^3 a_o)}^{1\over2}}
\eqno(13b).$$  

This is the expression derived by Felten (1984) written in
a somewhat different form.  At first sight, it may seem odd to
use terms such as $\Omega_o$ which are valid for Friedmann
cosmology but have no obvious relevance to MOND cosmology.  But it is
proven below that MOND on small scale is consistent with Friedmann
on large scale.

\section{A critical length scale for Modified Dynamics}

I now consider the meaning of the initial radius $r_i$ in eqs.\ 12
and 13.  Looking
back at the  Newtonian expression eq.\ 3 we see that, 
at any epoch characterized
by some value of density and pressure, the acceleration of the 
radius of the shell increases linearly with r.  This implies that there
should be some critical radius $r_c$, beyond which the acceleration
exceeds the MOND acceleration $a_o$ and the dynamics is Newtonian.  
That is to say, on all scales
greater than $r_c$, the usual Newtonian expressions, eqs.\ 3 and 5
apply to the expansion of a spherical region and the evolution of
the scale factor.

This critical length scale is given by $$r_c = \sqrt{{GM}\over a_o}
\eqno(14)$$ where again M is the active gravitational mass.
Making use of eqs.\ 4-7, eq.\ 14 becomes
$$r_c = a_o\Bigl|{{{\Omega_o{H_o}^2}\over{2x^3}} + {{\Omega_r{H_o}^2}\over
{x^4}} - {\lambda{H_o}^2}}\Bigl|^{-1} \eqno(15a)$$
or
$$r_c = {{2a_o}\over{\Omega_o{H_o}^2}}x^3 \eqno(15b)$$
when the universe is matter dominated, 
$$r_c = {{a_o}\over{\Omega_r{H_o}^2}}x^4 \eqno(15c)$$
when the universe is radiation dominated, and
$$r_c = {{a_o}\over{\lambda{H_o}^2}} \eqno(15d)$$ when the universe is
vacuum energy dominated.  Therefore, at any epoch 
characterized by a scale factor x, modified dynamics can only apply
in regions smaller than $r_c$.  This critical radius grows faster than
both the scale factor and the horizon, $l_h$, in the
radiation- and matter-dominated regimes (i.e., $r_c/l_h \propto t^2$).  

In Fig.\ 1 $r_c$ is plotted against scale factor.  Here and elsewhere
below the cosmological parameters are taken to be $H_o = 75$ km/(s Mpc),
$\Omega_o = 0.02$, $\Omega_r = 4.48\times 10^{-5}$ and $a_o = 1.2\times
10^{-8}$ cm/s$^2$.  This value of $\Omega_o$ is consistent with
the baryonic content of the Universe 
implied by considerations of primordial nucleosynthesis 
(Walker et al. 1991) and with estimates of the stellar mass in
galaxies and intra-cluster hot gas (Carlberg et al. 1998);  in the
context of MOND this would be the total matter content of the Universe
(i.e., no significant contribution from non-baryonic dark matter).
The density parameter in radiation, $\Omega_r$, is exactly that for
a black body of 2.73 K and the assumed Hubble parameter.  The value of
the acceleration
parameter, $a_o$, is that determined from fitting to the extended
rotation curves of nearby galaxies (Begeman et al. 1990). For the purposes
of this plot the cosmological constant has been set to zero.  The age
of a model universe with this assumed H$_o$ and
$\Omega_o$ would be $1.26 \times 10^{10}$ years which is consistent
with the recent determinations of the ages of globular clusters in
light of the new cluster distance scale (Chaboyer et al. 1997)

Also shown in Fig.\ 1 is the horizon scale ($\approx ct$) as a function
of scale factor determined by numerical integration of eq.\ 5.
It is evident that at early epochs $r_c$ 
is much smaller than the horizon size. This suggests that any causally
connected region of the Universe 
can be isotropic and homogeneous with the expansion governed by
the usual Friedmann equation, eq.\ 5.  That is to say, for
spatial separations larger $r_c$, it is valid to apply a Universal scale 
factor and, presumably, the RW metric.
However, about a typical point in the Universe there exist a smaller volume 
with radius $r_c$ within which modified dynamics and eq.\ 12 applies;
i.e., in which separations cannot be described in terms of a scale factor and
which expand at a slower rate than the universe at large.  Thus
at any epoch inhomogeneities must be present on a scales smaller than $r_c$.
For a universal scale factor greater than 0.23,
the critical MOND radius $r_c$
exceeds the horizon scale.  This means that the entire causally connected
Universe has become MONDIAN at a redshift of about 3.3 and can no longer
be described by the RW metric.  

This interpretation does contain a logical conundrum.  
In an actual MOND universe with Friedmann expansion
on a horizon scale but slower MOND expansion on sub-horizon scales,
not every point in the fluid can possibly be a center of 
MOND-dominated expansion and collapse;
it is not possible that a horizon-scale volume can expand while
smaller spherical regions about every point within that volume re-collapse.
This is a paradox of the present non-relativistic treatment which
can only be resolved in a more complete theory-- a theory involving
fluctuations in which local peaks probably play the role of seeds for
MOND-dominated expansion and re-collapse with voids developing elsewhere.
But, in any case, is likely that $r_c$ is the approximate scale below which
there exist MOND-induced inhomogeneities at any epoch in an evolving Universe.

Accepting this interpretation, we note that larger and larger masses
enter the MOND regime as the universe evolves.  Given that $r_o$ is the 
comoving scale of mass  $M_c$  and taking $x = r_c/r_o$, we have
$$M_c = {{\Omega_o{H_o}^2}\over {2G}}\Bigl({{r_c}\over {x}}\Bigr)^3
\eqno(16)$$  
This is also shown in Fig.\ 1 where
it is evident that objects of galaxy mass ($10^{11}$ M$_\odot$)
enter the MOND regime
at $x \approx 7\times 10^{-3}$ corresponding to a redshift of 140. 

Obviously then the appropriate value for the initial radius, $r_i$ in
eq.\ 13a, would be the radius at which eq.\ 12 first applies to the
expansion of the spherical region; i.e., $$r_i = r_c.\eqno(17)$$  
This would be about 14 kpc for the galaxy size region.  Moreover, the
initial expansion velocity would be given by the Hubble expansion
on a scale of $r_i$:
$${u_i} = Hr_i = r_iH_o\sqrt{\Omega_o(r_o/r_i)^3 + \Omega_r(r_o/r_i)^4
+(1-\Omega_o)(r_o/r_i)^2} \eqno(18)$$ which is about 320 km/s for the
the $10^{11}$ M$_\odot$ region.  We then find,
from eq.\ 13, that $q^2 = 1.25$ and $r_m/r_i = 3.5$; i.e., 
after entering the MOND regime at a redshift in excess of 100,
a galaxy mass would only expand by a factor of about four before re-collapsing.
The time-scale for this expansion is given by 
$$\Delta t = \sqrt{\pi}q{H^{-1}}\,\Bigl({{r_m}\over{r_i}}\Bigr)
\,{erf(q)}\eqno(19)$$ (Felten 1984); i.e.,
objects of galaxy size and smaller enter the MOND regime early and fall
out of Hubble expansion
on a time-scale comparable to the age of the Universe at that epoch.  For
the $10^{11}$ M$_\odot$ sphere this would be approximately 
$3\times 10^8$ years.
Re-collapse and virialization might take three or four times longer,
so we see that with modified dynamics in the context of a low density
Friedmann universe, galaxies should be in place 
as virialized objects when the Universe is about $10^9$ years old 
corresponding to a redshift of 9 or 10.

\section{The early radiation dominated Universe}

In the early universe the size of spherical regions dominated by modified
dynamics is small compared to the horizon.  
Taking eq.\ 15c and noting that the temperature of
the black body radiation scales with the inverse of the scale factor, we
have $$r_c = {{a_o}\over{\Omega_r{H_o}^2}}\Bigl({{T_o}\over T}\Bigr)^4 
\eqno(20)$$ 
which, for the assumed cosmology becomes
$$r_c = 4.54\times 10^{31}\Bigl({{T_o}\over T}\Bigr)^4\,\,cm\eqno(21)$$
With $T_o = 2.73$ K and $T=10^9$ K, corresponding to the epoch of
nucleosynthesis, we find $r_c = 2.5\times 10^{-3}$ cm.
The scale of the horizon in the radiation-dominated era is given by 
$$l_h \approx 0.5{(\Omega_r{H_o}^2})^{-{1\over 2}}(\Bigl({{T_o}\over
T}\Bigr)^2 c \eqno(22)$$
which is $10^{13}$ cm at the epoch of nucleosynthesis.
That is to say, the expansion of the Universe as a whole is identical
to that of the standard Big Bang with the scale of modified dynamics 
being 15 orders of magnitude smaller than the horizon size.

But because nucleosynthesis occurs on a smaller scale still-- that
of internucleon spacing ($\approx 10^{-7}$ cm)-- 
we must consider the fate of these small regions
of non-standard dynamics.  Taking eqs.\ 4 and 9 but now with 
the equation of state for radiation, we find
$$\ddot {r} = -(\Omega_r{H_o}^2a_o{r_o}^4)^{1\over 2}r^{-{3\over 2}}
\eqno(23)$$ which integrates to
$${\dot{r}}^2 = {u_i}^2 - 4(\Omega_r{H_o}^2{a_o}{r_o}^4)^{1\over 2}
\Bigl[{1\over{{r_i}^{1\over 2}}} - {1\over{r^{1\over 2}}}\Bigr]\eqno(24a)$$
Given eq.\ 18 for $u_i$ and that $x = r_c/r_o$
we find with eq.\ 15c that
$${u_i}^2 = (\Omega_r{H_o}^2{a_o}^2{r_o}^4)^{1\over 3}.\eqno(24b)$$
Then it is straight forward to show that, in the absence of other
considerations, any such region will expand more slowly than the
universe as a whole and will re-collapse after expanding by a factor 
$r_m/r_i = 16/9$.

This might well seem devastating not just for primordial nucleosynthesis, 
but also for the overall homogeneity of the early universe.  There
is, however, another physical effect which must be considered.  The slower 
expansion of these regions on the scale of $r_c$ will very rapidly lead to
density and hence pressure gradients which will resist re-collapse.  
Because the gravitational acceleration in these regions is so small,
only small density gradients are required
to keep these regions expanding with the Universe at large.
The gravitational acceleration in these
small MOND regions is (by definition) on the order of $a_o$.  So the
pressure gradient required to resist re-collapse can be estimated from
$${1\over\rho}{{dp}\over {dr}} = a_o\eqno(25)$$
Setting $p = {1\over 3} \rho c^2$ and $dr = r_c$ and making use of eq.\ 20
we estimate the corresponding density fluctuation over this scale to be 
$${{\delta \rho}\over {\rho}} = {{3{a_o}^2}\over{\Omega_r{H_o}^2c^2}}
\Bigl({{T_o}\over T}\Bigr)^4
\eqno(26)$$ which implies ${\delta\rho}/{\rho} \approx 10^{-31}$ when
$T=10^9$ K.  In 
the early radiation dominated universe, modified dynamics results in no
significant deviation from homogeneity and the thermal history is identical
to that of the standard Big Bang.  This means that all of the results
on nucleosynthesis in the standard model carry over to a MOND cosmology.

At the epoch of recombination (T = 4000 K), where radiation still 
dominates in low $\Omega_o$ models, we find $\delta \rho/\rho \approx
4\times 10^{-10}$.  That is to say, the MOND-induced inhomogeneities
would be five orders of magnitude less than the density fluctuations
implied by the observed fluctuations in the CMB.

The argument on pressure gradients resisting MOND collapse can be
re-framed in terms of the critical Jeans mass for gravitational 
instability.  The Jeans mass is, effectively, identical to the virial 
mass which, in the context of MOND, is given by
$$M_J = {9\over{Ga_o}}{c_s}^4\eqno(27)$$
where $c_s$ is the sound speed in the fluid being considered (Milgrom 1989).  
Before decoupling of matter and radiation, $c_s = c/\sqrt{3}$ 
implying that $M_J\approx c^4/Ga_o$ which, given that $a_o\approx cH_o$,
is on the order of the total mass of the present observable Universe.
Obviously this is vastly greater than the mass enclosed
in a MOND region (at the epoch of nucleosynthesis this is 
approximately $10^{-12}$ g); MOND dominated gravitational collapse is
clearly an impossibility before hydrogen recombination.

After recombination, the Jeans mass of the baryonic component becomes
$$M_J = {9\over{Ga_o}}{\Bigl({{kT_m}\over m}\Bigr)}^2 \eqno(28)$$
where $k$ is the Boltzmann constant, $m$ is the mean atomic mass,
and $T_m$ is the temperature of the matter.  But collapse can still 
not occur until the Jeans mass falls below the critical mass 
in a MOND dominated region.  From eqs.\ 16 and 20 this is found to be 
$$M_c = {{{a_o}^3\Omega_o{H_o}^2}\over{2G(\Omega_r{H_o}^2)^3}}
{\Bigl({{T_o}\over T}\Bigr)}^9 \eqno(29)$$
in the radiation-dominated regime.
We see from eqs. 28 and 29 that while the Jeans mass decreases with
the square of the radiation temperature, the critical MOND mass
rapidly increases as the temperature falls.  This is illustrated in
Fig.\ 2 after the epoch of recombination ($T_{rec}\approx 4000$ K) assuming
that $T_m = T^2/T_{rec}$ (true for non-relativistic mono-atomic fluid).
The MOND critical mass becomes comparable to the Jeans mass 
(eqs. 27 and 28) when the radiation temperature has fallen to 
$$T = \Bigl[{1\over{18}}{{\Omega_o{H_o}^2}
\over{(\Omega_r{H_o}^2)^3}}
\Bigl({m\over{k}}\Bigr)^2{a_o}^4{T_o}^9{T_{rec}}^2
\Bigr]^{1\over{13}}; \eqno(30)$$ 
with the cosmology 
assumed above this is $2.5\times 10^3$ K, or somewhat later
than the epoch of recombination (this value is obviously quite
insensitive to the actual values of the cosmological parameters).  
The corresponding value of the Jeans mass and critical MOND mass
is about $10^5$ M$_\odot$.  This means that shortly after 
recombination pressure gradients are no longer effective in preventing
MOND-induced collapse.  However, it is argued below that
$M_J<M_c$ is a necessary
but not a sufficient condition for MOND-dominated expansion and collapse.

\section{The formation of structure}

Expansion of a low density universe ($\Omega_o<<1$)
will remain radiation-dominated until
well after recombination;  for the cosmology assumed here this occurs
at $x = 4.48\times 10^{-3}$ ($z=222$).  It is clear from 
Fig.\ 2 that regions having a
mass less than $4\times 10^9$ M$_\odot$ enter the MOND regime while radiation
still dominates the energy density of the Universe.
Taken at face value then, it would seem that one should apply eq.\ 24 to
the MOND-dominated expansion of regions above the Jeans mass
which enter the MOND regime in the period between 
recombination and matter-radiation density equality.
However, during this period the horizon is much larger than the
scale over which MOND applies (at matter-radiation equality the horizon
is 4 Mpc but the scale of modified dynamics is only 3 kpc).
After recombination,
the photons are uncoupled to the matter, so it would be impossible for a
MOND region to re-collapse while the passive gravitational mass in
radiation still dominates the gravitational
deceleration;  the photons free-stream to the horizon.
MOND-dominated expansion and re-collapse as described by eq.\ 12 
would begin for all masses between 300 M$_\odot$ (the Jean's mass)
and $4\times10^9$ M$_\odot$ (the critical MOND mass)
at the epoch of matter-radiation equality (indicated by the heavy solid
line in Fig.\ 2).

The actual mass in a MOND-dominated region at the epoch of matter-radiation
equality, $M_e$, is extremely sensitive to the cosmological parameters.
This may be determined from eqs.\ 15 and 16 
(setting 15b equal to 15c) and is found to be 
$$ M_e = {{32{a_o}^3{\Omega_r}^6}\over{G{\Omega_o}^8{H_o}^4}} \eqno(31)$$
Combining with eq.\ 7 we find
$$M_e = {3.7\times 10^9}{\Bigl[\Bigl({{\Omega_o}\over{0.02}}\Bigr)
\Bigl({{H_o}\over{75}}\Bigr)^2\Bigr]}^8\,\,{\rm M_\odot}\eqno(32)$$
Matching the observed light element abundances with the
predictions of primordial nucleosynthesis (Walker et al. 1991), 
implies that $0.018<\Omega_o{(H_o/75)}^2<0.027$.
Then with eq.\ 32 we find that 
$4.3\times 10^8\,M_\odot<M_e< 3.7\times 10^{10}\, M_\odot$.

It is interesting that the mass scale over which MOND applies at the
epoch of matter-radiation equality-- when MOND collapse can begin and
significant inhomogeneities can form-- corresponds to that of low to
moderate mass
galaxies.  Perhaps this offers some explanation for the fact that the
lowest-mass virialized building-blocks of the Universe are galaxies
(the existing globular clusters and dwarf galaxies
re-collapsed simultaneously and may have survived due to incomplete 
merging).  But in any case, for objects of any mass scale, the separation of
from the Hubble expansion and subsequent re-collapse can be described
by eq.\ 12 (the Felten equation) with the initial radius being the critical
MOND radius given by eq.\ 15 for objects of mass greater than $10^9$ 
M$_\odot$, and the scaled comoving radius ($r = xr_o$) at the epoch
of matter-radiation equality for lower mass regions.

This dynamical history is shown for various mass scales in Fig.\ 3 which
is a plot of the scale factor ($r/r_o$) of regions of different mass
as a function of cosmic time to the point of maximum expansion.  
These curves are determined from numerical integrations
of eq.\ 12.  The cosmic scale factor corresponding to
Friedmann expansion (eq.\ 5) is also shown (again the cosmological 
term has been set to zero).  The vertical dotted line shows the epoch
of matter-radiation equality for this particular cosmological model.

It is evident that objects of globular cluster mass ($10^5$ M$_\odot$)
re-collapse very soon after matter-radiation equality;  maximum expansion is
reached at a cosmic time of $2.3\times 10^7$ years corresponding to a
redshift of 156.  Massive galaxies ($10^{11}$ M$_\odot$) reach this point of
maximum expansion at $t = 3\times 10^8$ years or $z=26$. 
Clusters of galaxies ($10^{14}$ M$_\odot$) begin to re-collapse at
$2.65\times 10^9$ years (z=3).  

The mass which is just turning
around at the present epoch is $3.7\times 10^{15}$ M$_\odot$.
The comoving scale is 66 Mpc but the present radius would be 29 Mpc.
This would correspond to the mass and scale of superclusters as noted
by Felten (1984).  Taken literally the implication would be that a region of 
30 Mpc about a typical observer should be collapsing rather than 
expanding, which is evidently not the case locally.  This result, which
might be taken as an argument against a pure MOND cosmology, neglects
likely complications arising in a real Universe filled with significant
density enhancements and peculiar accelerations.
In the fully-developed MOND universe at the present epoch, 
the large scale inhomogeneities and
resulting tides will most likely cause large aspherical distortions of 
the developing structure.  Thus, the effects of distant matter cannot
be ignored; i.e., the fundamental assumption underlying this
treatment of isolated regions breaks down.  Given the enhanced
tidal effects and the fact that, in MOND, the internal dynamics of a 
a region is effected by the external acceleration field, the 
``external field effect'' (Milgrom 1983a),  the growth of pancakes, filaments
and voids would seem natural.  The present turnaround
radius of 30 Mpc may only give an estimate of the scale of structure
which has significantly separated out of the present Hubble flow, as 
suggested by Felten (1984).

Since the entire present Universe is MONDIAN (in the absence of a 
dynamically significant
cosmological constant), then on all scales out to the horizon, the expansion 
cannot be uniform; the mean value of the
Hubble parameter grows with scale.  This also implies that the mean
density of matter within a spherical region should decrease out to the 
horizon.  This may be determined by integrating eq.\ 12 for spheres with
comoving radii larger than 66 Mpc (corresponding to the current shell
which is just turning around at 29 Mpc) out to the horizon scale of
4000 Mpc.  The result is shown in Fig.\ 4 which is a log-log 
plot of the ratio of
the mean density inside a finite spherical volume to the mean 
density of the Universe as a function the present radius of the sphere
(not the comoving radius).  It is evident that the average density
smoothly increases from the horizon down to a scale of 30 Mpc where it is
10 times larger than the mean density.  

The literal and naive interpretation of Fig.\ 4, and that which would seem
most consistent
with the treatment of isolated spherical regions, is that this would 
represent the density distribution about a single observer in
in a MOND Universe.  But if we require, consistent with the Cosmological
Principle, that the observer have no special position, then an equally
valid interpretation is that the average density distribution
about any observer in the MOND universe declines smoothly to the
horizon-- that the matter distribution is
non-analytic (fractal) and does not imply a special position for one observer
(Coleman \& Pietronero 1991).

It has been claimed, from analysis of redshift catalogues, that the mean 
density of galaxies about a given galaxy
does decrease with scale (eg. Coleman \& Pietronero 1991) out to 
cosmological distances.  Although this claim is controversial (Peebles 1993), 
it is generally consistent with expectations for a pure MOND cosmology.  
The actual radius of cross-over to homogeneity would depend upon the
value of a possible cosmological constant.
With a cosmological term large enough to be dynamically significant
($\lambda\approx 1$), 
the critical radius for modified dynamics (eq.\ 15a), after
first becoming infinitely large, asymptotically approaches a constant value: 
$r_c\rightarrow (1/6)(c/H_o)$, i.e., about 1/6 the current horizon
if $H_o\approx$ 75 km/s-Mpc (the observed value of
$a_o$ would then correspond to $cH_o/6$).    
In this case we might expect large scale homogeneity and uniform exponential
expansion of the Universe on sub-horizon scales larger than several
hundred Mpc.

\section{Conclusions} 

Although there is not yet a plausible candidate for a 
general theory of gravity which predicts the MOND phenomenology in the
limit of low accelerations, consideration of the dynamics of a finite
spherical volume may contain elements of a realistic MOND cosmology.
Perhaps the most interesting conclusion that can be drawn from such
a exercise is that modified dynamics over finite separations
is compatible with Friedmann
cosmology on large scale; until relatively recently in the history
of the Universe MOND could  
not dominate the dynamical evolution of the Universe in general.  At
earlier epochs, the scale over which MOND applies, 
$r_c$ (within which the deceleration of Hubble expansion is less 
than $a_o$), is smaller 
than the size of a causally connected region which implies that the 
Universe as a whole can be isotropic and adequately described by the 
RW metric with expansion governed by the usual Friedmann equation.
However, the fact that the expansion is slower in MOND-dominated regions
implies that inhomogeneities must be present at any epoch on a scale
of $r_c$ and smaller.  In the 
early radiation dominated Universe, the magnitude of these 
MOND-induced inhomogeneities
is very small ($\delta \rho/\rho \approx 10^{-31}$ when T=$10^9$ K),
because of the small pressure gradients required to restore uniform
expansion.  Therefore, the thermal and dynamical history of the early
MOND Universe is exactly that of the standard Big Bang and all predictions
relevant to the nucleosynthesis of the light elements carry over to
MOND cosmology.

After non-relativistic matter dominates the mass density of the Universe
(which can be rather late in a low density Universe), MOND cosmology 
diverges from that of standard cosmology.  At the epoch of matter-radiation
equality, objects with mass up to $4\times 10^9$ M$_\odot$ rapidly collapse 
to form virialized objects.  The fact that this mass scale, which is the 
mass in the MOND regime at radiation-matter equality, is comparable to that 
of low mass galaxies seems significant:  objects of this mass would 
be the principal virialized building blocks in the Universe.
Moreover, this mass scale emerges naturally from the basic dynamics;
astrophysical considerations such as cooling vs. collapse time scales
do not play a role.
Although objects of smaller mass (down to $10^2$ 
M$_\odot$) collapse and virialize first, a process probably
accompanied by star formation, these objects would rapidly merge
in the the larger collapsing regions.  This 
suggests that galaxy formation is primarily dissipationless; that the
stellar content of galaxies may be in place before the galaxies actually
form.  Of course, early star formation 
could be limited by processes such as photo-dissociation of H$_2$ as in
standard scenarios (Haiman et al. 1997); these self-limiting processes 
could keep much of
the matter content of the universe in gaseous form as seems to be implied by
the observations of rich clusters.

Many of these low mass galaxies galaxies would merge to form 
more massive objects as
larger and larger scales come into the MOND regime.  
A spherical region with the mass of a large galaxy 
($10^{11}$ M$_\odot$) reaches maximum expansion and begins to
re-collapse at a redshift of 26 which implies that
large galaxies should be in place as virialized
objects by redshift of 5 to 10.  This is earlier than the
epoch of galaxy formation in the standard CDM paradigm (Frenk et al. 1988).
Moreover, from Fig.\ 3 it is evident that regions with the size and
mass of a cluster have reached maximum expansion by a cosmic age of
$2.7\times 10^9$ years corresponding to a redshift of three.  This means
that by z=3 not only do massive galaxies exist but they are also 
significantly clustered (the density of the $10^{14}$ M$_\odot$ region
would be enhanced by a factor of 6.5 over the mean at this redshift).
This may be relevant to the observation of luminous galaxies at z=3
which is remarkable not only because they are there but also because
of the apparent degree of clustering (Steidel et al. 1997).  Such
observations may be able to distinguish between the cosmogony sketched
here and that of the standard CDM paradigm.

The largest objects being virialized
now would be clusters of galaxies with masses in excess of $10^{14}$
M$_\odot$.
Superclusters would only now be reaching maximum 
expansion.
Such a scenario of structure formation is hierarchical in the extreme
and as such bears a resemblance to the more standard (CDM) scenarios
of the build up of structure.  But here, dissipationless 
dark matter is not required to enhance structure formation;  structure 
forms inevitably on the MOND scale of $r_c$ because of the effective
logarithmic potential.  

In the present Universe regions approaching the horizon scale
would be subject
to a scale-dependent deceleration due to 
modified dynamics.  This would lead to a Universe in which the mean run
of density about a galaxy 
decreases smoothly to a cosmic scale.  The actual scale for approach to
homogeneity would depend upon the value of the cosmological constant;
for $\lambda \approx 1$ corresponding to a zero curvature Universe, the 
density and expansion of the Universe would be more or less uniform on 
scales greater than several hundred Mpc.  In such a Universe, the Hubble 
parameter would also be, on average, scale dependent, increasing with the
separation between objects out to some significant fraction of the Hubble
radius.  However, these are the aspects of MOND 
cosmology which most depend
upon the unknown properties of the underlying theory-- such as the 
value of the cosmological constant and whether or not MOND phenomenology 
saturates at some lower value of acceleration and attraction 
returns to inverse square.

The details of cosmology and cosmogony sketched here
are dependent upon the assumptions which underly this procedure:
no effects of the surrounding universe on the finite spherical
volume, no return to Newtonian dynamics at lower accelerations,
no variation of $a_o$ with cosmic time.
A different set of assumptions is also plausible and would lead
to a different MOND cosmology.  The remarkable aspect of the cosmology
resulting from these assumptions 
is the fact that the pre-recombination dynamical and thermal evolution
is identical to that of the standard Big Bang.
But in any case, it seems inevitable that in a MOND cosmology, 
structure formation proceeds
much more rapidly and efficiently than in standard cosmologies due to 
the effective logarithmic potential.  

Friedman expansion on horizon scale combined with modified dynamics
on smaller scale suggests that density peaks may be required to
play the role of seeds or centers about which 
MOND-dominated expansion and re-collapse occurs.
Apart from this, primordial density fluctuations have played no role in 
the present discussion of structure formation.
In the correct relativistic theory this will probably not be the case;  
for example, in
stratified aquadratic scalar-tensor theory (Sanders 1997), there are no
effects of modified dynamics in the absence of scalar field gradients;
in a perfectly homogeneous Universe, the metric is RW and structure
never develops.  This suggests that in a proper theory fluctuations may
be essential for the development of structure which may then proceed,
qualitatively, as described above.  

One should be cautious about pushing these results too far.
In standard theory, the Newtonian dynamics of an expanding region takes on
cosmological significance only in retrospect, that is, after the application
of General Relativity and the construction of a relativistic cosmology. 
Here, the order is reversed-- the rules of MOND are applied to a finite
spherical region before the development of the appropriate relativistic 
theory.  But it is possible that  
many of the aspects of a fully relativistic cosmology may be previewed, 
at least in a qualitative sense, by such an exercise.
While this remains to be seen, it is of considerable interest that
the resulting cosmology does seem  
to reconcile the extreme homogeneity of the early radiation-dominated
Universe with early galaxy formation and the extreme range of structure 
observed in the matter-dominated era.  Moreover, the fact 
that this can be accomplished without the necessity of 
invoking hypothetical non-baryonic dark matter is entirely consistent with 
the original motivation for modified dynamics as an alternative to dark 
matter on the scale of galaxies and galaxy clusters.

\acknowledgments

I am very grateful to J.D. Bekenstein and M. Milgrom for helpful comments
on this work.

\clearpage

\figcaption[ ] {A log-log plot of the evolution of the horizon, $l_h$ 
(solid line), and
the radius within which MOND applies, $r_c$ (dotted line), both in Mpc, as a
function of cosmic scale factor for the given cosmological parameters
(H$_o$ = 75 km/s-Mpc, $\Omega_o = 0.02$). Also shown is the 
mass enclosed within $r_c$ (dashed line) in units of $10^{11}$ M$_\odot$.
The vertical long dashed line indicates the epoch at which matter
and radiation contribute equally to the deceleration.  It is evident that
when the scale factor is smaller than 0.23 ($z > 3.3$) the region over which
modified dynamics applies is smaller than the size of a causally connected
region. \label{fig1}}

\figcaption[ ] {A log-log plot of the critical mass within a MOND-dominated
(solid line) and the Jeans mass in the context of MOND
(dotted line), both in M$_\odot$, as a function of cosmic scale factor,
for the adopted cosmological parameters.
The epochs of recombination and matter-radiation equality are shown by
the vertical dashed lines.  MOND-induced collapse proceeds rapidly after
matter-radiation equality with initially all masses between the Jeans
mass (334 M$\odot$) and the critical MOND mass ($3.7\times 10^9$ M$_\odot$)
separating out of the Hubble flow and re-collapsing.  This range is indicated
by the heavier solid line. \label{fig2}}

\figcaption[ ] {A log-log plot of the scale factor for regions of various
mass (the radius over the comoving radius) as a function of cosmic time, for
the adopted cosmological parameters.
The MOND cosmogony hierarchical and bottom-up, with 
small structures forming soon after matter-radiation equality (indicated by
the vertical dotted line).  Massive galaxies ($10^{11}$ M$_\odot$) separate
out of the Hubble flow early and reach maximum expansion at a cosmic age
of $3\times 10^8$ years.  Supercluster mass structures ($10^{16}$
M$_\odot$) are only now reaching maximum expansion and turn-around.
\label{fig3}}

\figcaption[ ] {A log-log plot of the mean density,
at the present epoch, inside spherical
shells as a function of radius in Mpc.  The density is given
in terms of the mean universal density.  It is evident that the density
decreases smoothly to the horizon implying that a MOND-dominated Universe.
at present, would be very inhomogeneous.  The scale for approach to
homogeneity is smaller if the cosmological constant is not zero
($\approx$ 600 Mpc for $\lambda\approx 1$).  \label{fig4}}

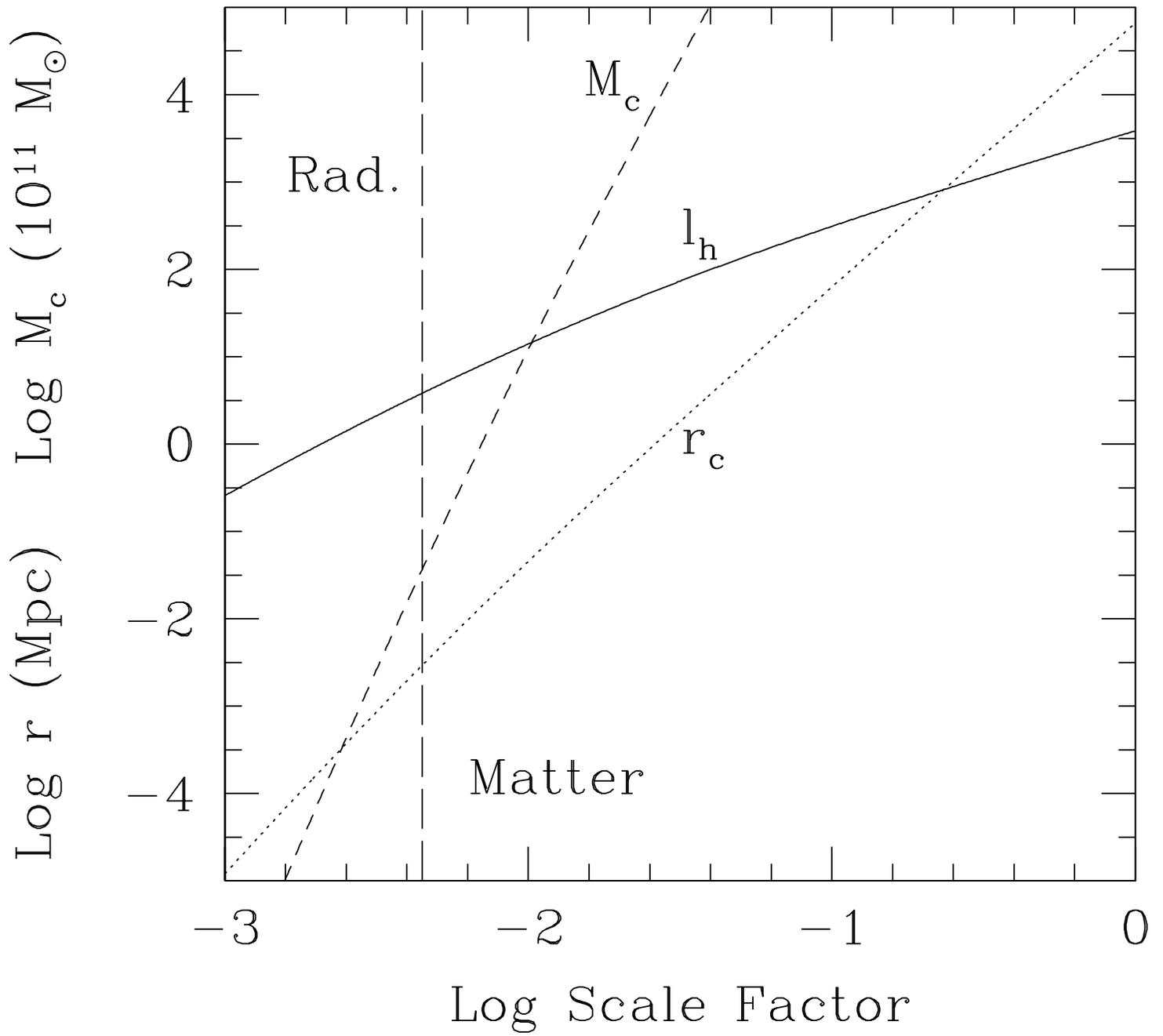
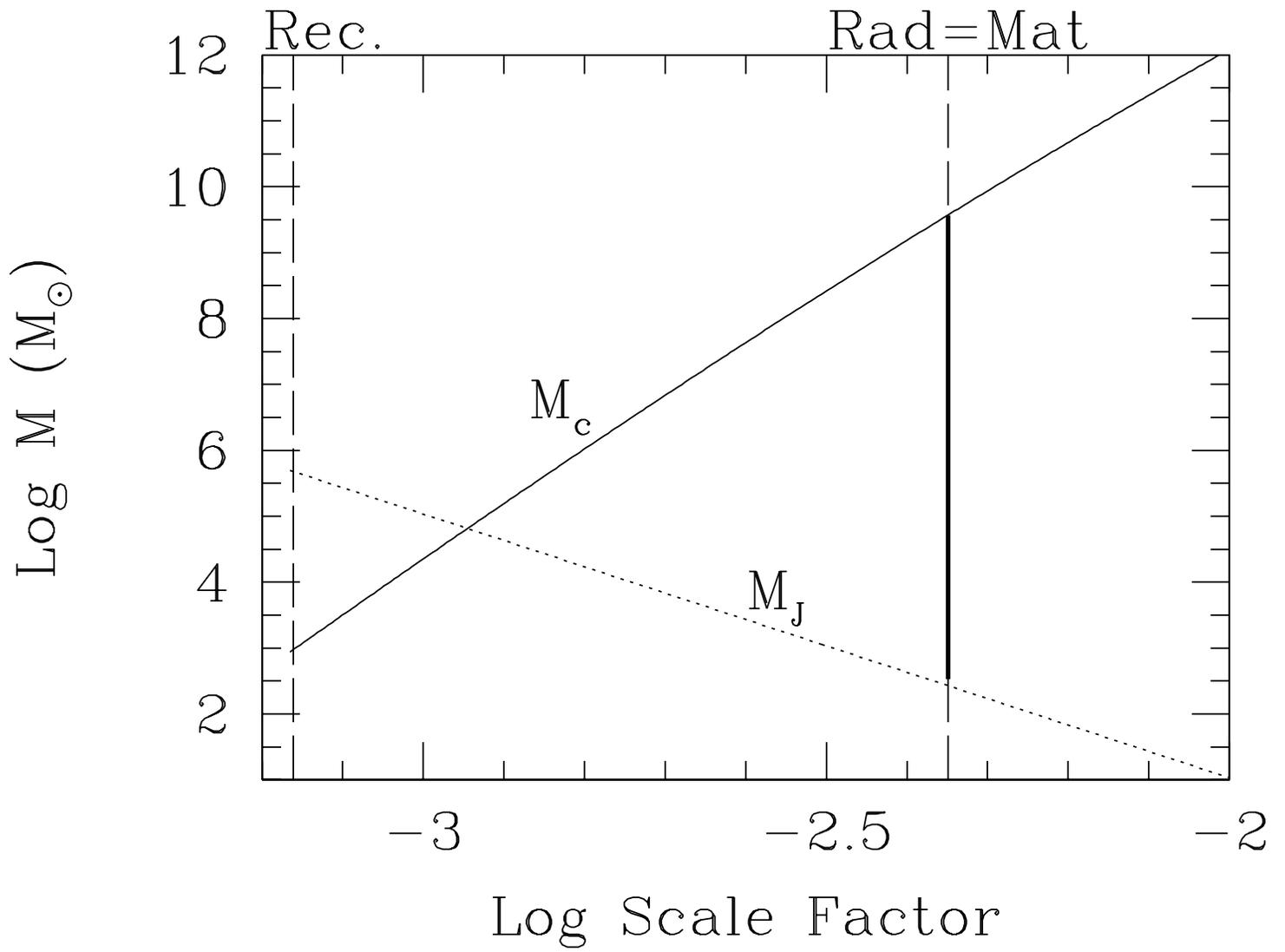
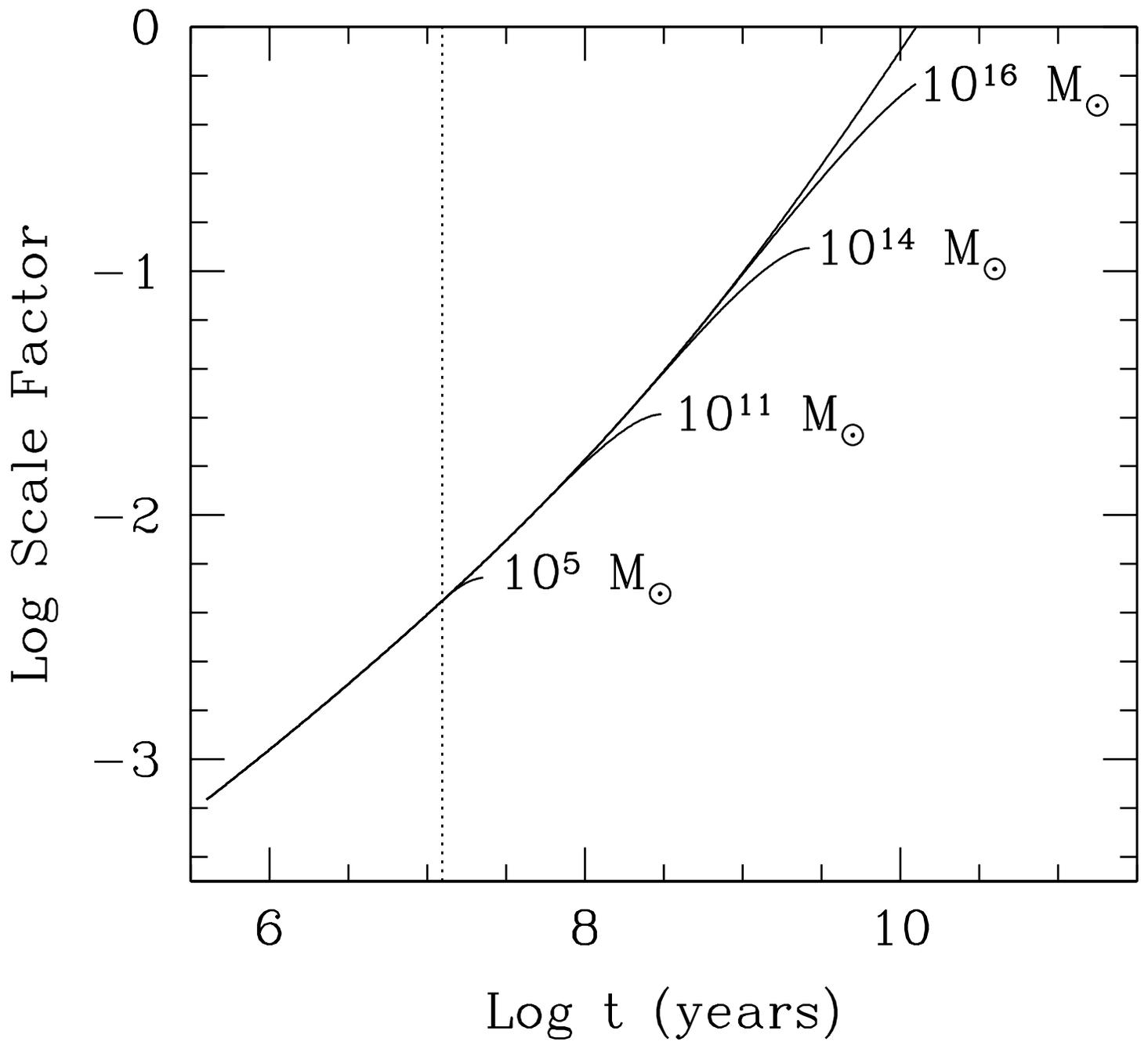
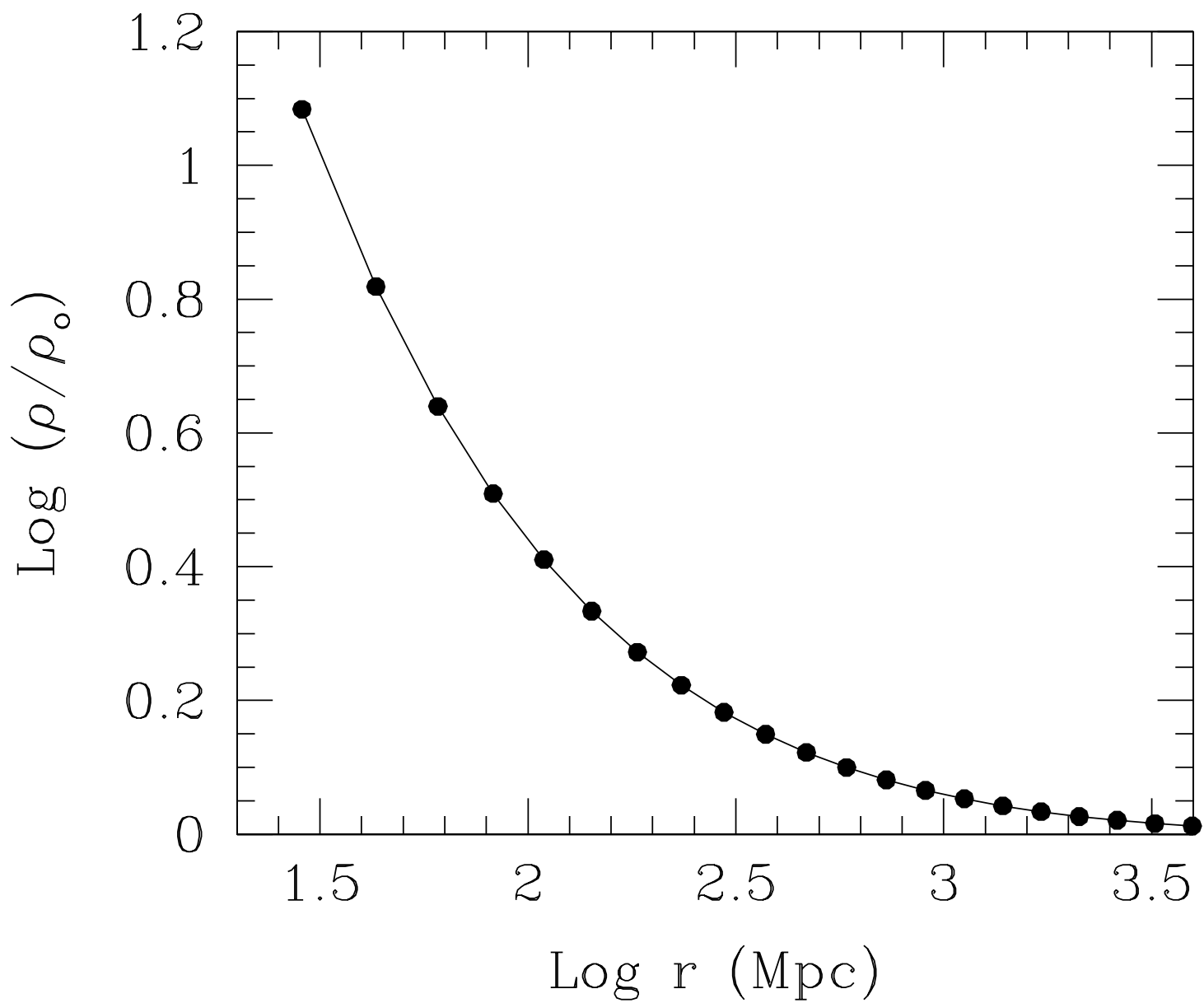
\end{document}